\documentclass[twocolumn,english]{revtex4-2}
\usepackage[T1]{fontenc}
\usepackage[latin9]{luainputenc}
\usepackage{amsmath}
\usepackage{babel}
\begin{document}
\title{The effect of Coulomb assisted hopping on STM signal: extended two
site Hubbard model analysis}
\author{Garry Goldstein}
\address{garrygoldsteinwinnipeg@gmail.com}
\begin{abstract}
In this work we study STM signal in the presence of Coulomb assisted
hopping. We perform an extended two site Hubbard model analysis between
the atom on the tip and the atom in the sample nearest to each other.
We show that in the presence of Coulomb assisted hopping the STM signal
depends on several spectral functions thereby complicating its interpretation.
Furthermore in the broadband tip limit there are now three different
competing rates for the total current (instead of one for the usual
two site Hubbard model analysis used in the literature so far). We
find an exact (within the Fermi golden rule - that is in the limit
of weak coupling between tip and sample) expression for the current
as a function of the bias voltage. As an example we apply our calculations
to the case of free fermions with a uniform density of states. Even
in this simple case there are non-trivial corrections - where the
$dI/d\mathcal{V}$ (the rate of change of the current with respect
to bias voltage) is not uniform as a two site (non-extended) Hubbard
model analysis would predict. We also show that for realistic conditions
the corrections predicted here are order one.
\end{abstract}
\maketitle

\section{\protect\label{sec:Introduction}Introduction}

Tunneling spectroscopy is a method which allows one to access the
spectral functions of the sample \citep{Coleman_2016,Chen_2021}.
Since its invention, by Binning et. al. in 1982 \citep{Binning_1982,Binning_1982(2)},
the Scanning Tunneling Microscope (STM), has revolutionized the field
of experimental surface and materials studies \citep{Chen_2021,Guentherodt_1994,Magonov_1996,Voigtlander_2015}.
At its core Scanning Tunneling Microscopy (STM) is based on the idea
that an atom of the tip is close enough to the sample that electron
tunneling is possible between the tip and the sample \citep{Chen_2021,Coleman_2016,Guentherodt_1994,Magonov_1996,Voigtlander_2015}
through the classically forbidden region. In a STM experiment the
probe tip is brought, using piezo-electronics, to within a fraction
of a nanometer of the sample surface \citep{Chen_2021,Guentherodt_1994,Magonov_1996,Voigtlander_2015}
allowing for tunneling. By applying a bias voltage between the tip
and the sample a current is generated as electrons tunnel from the
atom of the tip closest to the sample into the sample or in the opposite
direction depending on the sign of the bias voltage. The accomplishments
of STM experimenters have been numerous: 
\begin{itemize}
\item The first element sensitive imaging (of GaAs) by \citep{Freensta_1987,Freensta_1987(2)}
\item The introduction of the optical beam deflection methods by \citep{Meyer_1988}
\item The positioning of atoms on material surfaces with STM imaging by
\citep{Crommie_1993,Crommie_1993(2)}
\item Vibrational spectroscopy with STM was achieved by \citep{Stipe_1998,Stipe_1999}
\end{itemize}
to name a few. This research has so far been analyzed within the regular
(not extended) two site Hubbard model context for the two atoms on
the tip and sample closest to each other \citep{Coleman_2016,Chen_2021}.
The current as a function of bias voltage was obtained \citep{Chen_2021,Coleman_2016,Guentherodt_1994,Magonov_1996,Voigtlander_2015}
within the regular Hubbard model. Indeed most of the literature \citep{Chen_2021,Coleman_2016,Guentherodt_1994,Magonov_1996,Voigtlander_2015}
deals only with a Hubbard model analysis of the system where an onsite
Coulomb term $U$ is kept and a tunneling term $t$ is then added
- because of the overlap of the wave functions of the tip and the
sample - with no further two site extended Hubbard model terms considered
so far as far as the author is aware. Here we wish to extend this
STM analysis further by considering the extended two site Hubbard
model between the tip atom and the nearest sample atom \citep{Hubbard_1963,Hubbard_1964,Hubbard_1965,Fazekas_1999}.
We find that for the two site extended Hubbard model (with a single
relevant spinful orbital $\left\{ c_{\sigma}^{\dagger}\left(0\right),c_{\sigma}\left(0\right)\right\} $
for the sample atom and a single relevant spinful orbital $\left\{ \psi_{\sigma}^{\dagger}\left(0\right),\psi_{\sigma}\left(0\right)\right\} $
for the tip atom closest to it \citep{Fazekas_1999}) many additional
terms that lead to current between the tip and the sample. We compute
the current as a function of external voltage $\mathcal{V}$ in Eq.
(\ref{eq:Current-1}) in the broad band tip limit. The current now
depends on three unknown parameters $\Gamma_{1/2/3}$ (instead of
just one parameter as in previous STM analysis \citep{Chen_2021,Coleman_2016,Guentherodt_1994,Magonov_1996,Voigtlander_2015})
which are associated with properties of the tip and therefore are
hard to compute but can be included phenomenologically \citep{Chen_2021,Coleman_2016,Guentherodt_1994,Magonov_1996,Voigtlander_2015}.
As an application of this type of calculation we consider the case
when the sample is free fermions with a uniform density of states.
Already in that simple case there are non-trivial corrections to the
STM current (see Eq. (\ref{eq:Current_final}) where the $\frac{\partial I}{\partial\mathcal{V}}$
(the rate of change of bias current with respect to voltage) depends
explicitly on the bias voltage due to additional current terms we
computed. We also show that for realistic conditions the corrections
predicted here are order one. For simplicity, in our analysis, we
have neglected spin orbit coupling which is tedious but straightforward
to add.

\section{\protect\label{sec:Extended-Hubbard-model}Extended Hubbard model
review}

We consider the extended two site Hubbard model \citep{Fazekas_1999,Hubbard_1963,Hubbard_1964,Hubbard_1965}.
The Hamiltonian is well know and given by \citep{Fazekas_1999,Hubbard_1963,Hubbard_1964,Hubbard_1965}:

\begin{equation}
H_{pair}=H_{Hub}+H_{t}+H_{V}+H_{F}+H_{X}+H_{Y}\label{eq:Hubbard_Hamiltonian}
\end{equation}
Where 
\begin{align}
 & H_{Hub}=H_{\psi}\left(\left\{ \psi_{\sigma}^{\dagger}\left(0\right),\psi_{\sigma}\left(0\right)\right\} \right)+H_{c}\left(\left\{ c_{\sigma}^{\dagger}\left(0\right),c_{\sigma}\left(0\right)\right\} \right)\nonumber \\
 & H_{t}=\sum_{\sigma}\left[t\psi_{\sigma}^{\dagger}\left(0\right)c_{\sigma}\left(0\right)+t^{*}c_{\sigma}^{\dagger}\left(0\right)\psi_{\sigma}\left(0\right)\right]\nonumber \\
 & H_{V}=V\sum_{\sigma_{1}\sigma_{2}}\psi_{\sigma_{1}}^{\dagger}\left(0\right)\psi_{\sigma_{1}}\left(0\right)c_{\sigma_{2}}^{\dagger}\left(0\right)c_{\sigma_{2}}\left(0\right)\nonumber \\
 & H_{F}=-2F\left(\mathbf{S}_{\psi}\cdot\mathbf{S}_{c}-\frac{1}{4}n_{\psi}n_{c}\right)\nonumber \\
 & H_{\Omega}=\sum_{\sigma}\left[\Omega\psi_{\sigma}^{\dagger}\left(0\right)c_{\sigma}\left(0\right)n_{\psi\bar{\sigma}}\left(0\right)+\Omega^{*}c_{\sigma}^{\dagger}\left(0\right)\psi_{\sigma}\left(0\right)n_{\psi\bar{\sigma}}\left(0\right)\right]\nonumber \\
 & H_{\tilde{\Omega}}=\sum_{\sigma}\left[\tilde{\Omega}\psi_{\sigma}^{\dagger}\left(0\right)c_{\sigma}\left(0\right)n_{c\bar{\sigma}}\left(0\right)+\tilde{\Omega}^{*}c_{\sigma}^{\dagger}\left(0\right)\psi_{\sigma}\left(0\right)n_{c\bar{\sigma}}\left(0\right)\right]\nonumber \\
 & H_{Y}=Y\sum\left[\psi_{\uparrow}^{\dagger}\left(0\right)\psi_{\downarrow}^{\dagger}\left(0\right)c_{\downarrow}\left(0\right)c_{\uparrow}\left(0\right)+h.c.\right]\label{eq:Hamiltonians}
\end{align}
Here $H_{\psi}\left(\left\{ \psi_{\sigma}^{\dagger}\left(0\right),\psi_{\sigma}\left(0\right)\right\} \right)$
is the Hamiltonian for the tip and $H_{c}\left(\left\{ c_{\sigma}^{\dagger}\left(0\right),c_{\sigma}\left(0\right)\right\} \right)$
is the Hamiltonian for the sample and $\mathbf{S}$ and $n$ refer
to spin and electron number respectively. The expressions for $\Omega$,
$\tilde{\Omega}$, $t$, $V$, $F$ and $Y$ are well known in terms
of the single particle basis wave functions and the single particle
Hamiltonian and Coulomb interactions \citep{Fazekas_1999}. Indeed
we now introduce the following generating integral \citep{Fazekas_1999}: 
\begin{widetext}
\begin{equation}
V\left(c/\psi,c/\psi,c/\psi,c/\psi\right)=\int d^{d}\mathbf{r}_{1}d^{d}\mathbf{r}_{2}\varphi_{c/\psi}^{\ast}\left(\mathbf{r}_{1}\right)\varphi_{c/\psi}\left(\mathbf{r}_{1}\right)\frac{e^{2}}{\left|\mathbf{r}_{1}-\mathbf{r}_{2}\right|}\varphi_{c/\psi}^{\ast}\left(\mathbf{r}_{2}\right)\varphi_{c/\psi}\left(\mathbf{r}_{2}\right)\label{eq:Coulomb_integral}
\end{equation}
\end{widetext}

Where $\varphi_{c/\psi}$ is the single particle wavefunction for
the tip atom or the sample atom respectively (for this calculation
we are ignoring spin orbit coupling and just focusing on two sites
with a single orbital). Now we see that \citep{Fazekas_1999}
\begin{align}
V & =V\left(c,c,\psi,\psi\right)\nonumber \\
F & =V\left(c,\psi,\psi,c\right)\nonumber \\
\Omega & =V\left(\psi,c,c,c\right)\nonumber \\
\tilde{\Omega} & =V\left(\psi,\psi,\psi,c\right)\nonumber \\
Y & =V\left(\psi,c,\psi,c\right)\label{eq:Terms}
\end{align}
and $t$ is the usual tunneling \citep{Coleman_2016,Chen_2021}. Because
of the exponential suppression of tunneling with separation between
the particles (say within the WKB approximation \citep{Shankar_1994})
we have that 
\begin{equation}
F,Y\ll\Omega,\tilde{\Omega},t,V\label{eq:Small}
\end{equation}
as both involve the tunneling of two electrons through the forbidden
region between the sample and the tip and will be neglected (see \citep{Fazekas_1999}
and Eq. (\ref{eq:Terms})). Furthermore $H_{V}$ does not lead to
current between the sample and the tip, to leading order, and therefore
will also be dropped. In this work we will focus on the Hamiltonian
in Eq. (\ref{eq:Tunneling_Hamiltonian}) which has $H_{\Omega}$ and
$H_{\tilde{\Omega}}$ on top of $H_{t}$ (previously considered in
the literature \citep{Coleman_2016,Chen_2021}). 

\subsection{\protect\label{subsec:Estimate-of-ratio}Estimate of ratio between
$\Omega$, $\tilde{\Omega}$ and $t$}

We have that 
\begin{align}
t & =\int d^{d}\mathbf{r}\varphi_{\psi}^{\ast}\left(\mathbf{r}\right)\varphi_{c}\left(\mathbf{r}\right)V_{KS}\left(\mathbf{\mathbf{r}}\right)\nonumber \\
 & \cong\bar{V}_{KS}\int d^{d}\mathbf{r}\varphi_{\psi}^{\ast}\left(\mathbf{r}\right)\varphi_{c}\left(\mathbf{r}\right)\label{eq:Tunneling}
\end{align}
Here we have that $V_{KS}$ is the effective Khon Sham potential \citep{Martin_2020,Marx_2009,Singh_2006}
and $\bar{V}_{KS}$ is some typical value on the order of a few eV.
Whereas: 
\begin{align}
\tilde{\Omega} & =\int d^{d}\mathbf{r}_{1}d^{d}\mathbf{r}_{2}\varphi_{\psi}^{\ast}\left(\mathbf{r}_{1}\right)\varphi_{c}\left(\mathbf{r}_{1}\right)\frac{e^{2}}{\left|\mathbf{r}_{1}-\mathbf{r}_{2}\right|}\varphi_{c}^{\ast}\left(\mathbf{r}_{2}\right)\varphi_{c}\left(\mathbf{r}_{2}\right)\nonumber \\
 & \cong\int d^{d}\mathbf{r}_{1}d^{d}\mathbf{r}_{2}\varphi_{\psi}^{\ast}\left(\mathbf{r}_{1}\right)\varphi_{c}\left(\mathbf{r}_{1}\right)\frac{e^{2}}{\left|\mathbf{r}_{c}-\mathbf{r}_{\psi}\right|}\varphi_{c}^{\ast}\left(\mathbf{r}_{2}\right)\varphi_{c}\left(\mathbf{r}_{2}\right)\nonumber \\
 & =\frac{e^{2}}{\left|\mathbf{r}_{c}-\mathbf{r}_{\psi}\right|}\int d^{d}\mathbf{r}\varphi_{\psi}^{\ast}\left(\mathbf{r}\right)\varphi_{c}\left(\mathbf{r}\right)\label{eq:Coulomb_assisted}
\end{align}
 Here $\mathbf{r}_{\psi}$ and $\mathbf{r}_{c}$ are the positions
of the nuclei of the tip atom and the atom nearest to it. As such
we see that: 
\begin{equation}
\Rightarrow\frac{\tilde{\Omega}}{t}\cong\frac{\frac{e^{2}}{\left|\mathbf{r}_{c}-\mathbf{r}_{\psi}\right|}}{\bar{V}_{KS}}\label{eq:Ratio-1}
\end{equation}
 Now we put $\left|\mathbf{r}_{c}-\mathbf{r}_{\psi}\right|\cong0.5nm$
\citet{Chen_2021} so we obtain that $\frac{e^{2}}{\left|\mathbf{r}_{c}-\mathbf{r}_{\psi}\right|}\cong2.9eV$
or 
\begin{equation}
\frac{\Omega}{t}\sim\frac{\tilde{\Omega}}{t}\sim1\label{eq:One}
\end{equation}
As such all corrections are order one.

\section{\protect\label{sec:Green's-functions-review}Statement of the main
result}

Consider an STM tip in tunneling proximity to a sample. We have that
the Hamiltonian is given by \citep{Coleman_2016,Fazekas_1999}: 
\begin{align}
H= & H_{\psi}\left(\left\{ \psi_{\sigma}^{\dagger}\left(x\right),\psi_{\sigma}\left(x\right)\right\} \right)+H_{c}\left(\left\{ c_{\sigma}^{\dagger}\left(x\right),c_{\sigma}\left(x\right)\right\} \right)\nonumber \\
 & +H_{t}+H_{\Omega}+H_{\tilde{\Omega}}-e\mathcal{V}N_{\psi}\label{eq:Tunneling_Hamiltonian}
\end{align}
Here we have now included also the rest of the atoms of the tip and
the rest of the atoms of the sample. Here $\mathcal{V}$ is an external
applied voltage and $N_{\psi}$ is the total number of electrons in
the tip. We now introduce 
\begin{align}
G_{O_{1}O_{2}}^{R}\left(\omega\right) & =i\int d\tau\exp\left(i\omega t\right)\theta\left(\tau\right)\left\langle \left[O_{1}\left(\tau\right),O_{2}\left(0\right)\right]\right\rangle \nonumber \\
G_{O_{1}O_{2}}^{R}\left(\omega\right) & =i\int d\tau\exp\left(i\omega t\right)\theta\left(-\tau\right)\left\langle \left[O_{1}\left(\tau\right),O_{2}\left(0\right)\right]\right\rangle \nonumber \\
A_{O_{1}O_{2}}\left(\omega\right) & =-i\left[G_{O_{1}O_{2}}^{A}\left(\omega\right)-G_{O_{1}O_{2}}^{R}\left(\omega\right)\right]\label{eq:Greens_functions}
\end{align}
Here $A_{O_{1}O_{2}}\left(\omega\right)$ is the spectral function
for $O_{1}$ and $O_{2}$. We now introduce:
\begin{align}
O_{\psi}^{\downarrow} & =n_{\psi\downarrow}\left(0\right)\psi_{\uparrow}\left(0\right),O_{c}^{\downarrow}=n_{c\downarrow}\left(0\right)c_{\uparrow}\left(0\right)\nonumber \\
O_{\psi}^{\uparrow} & =n_{\psi\uparrow}\left(0\right)\psi_{\downarrow}\left(0\right),O_{c}^{\uparrow}=n_{c\uparrow}\left(0\right)c_{\downarrow}\left(0\right)\label{eq:Operators}
\end{align}
 Where $n$ is the density of electrons. Then the STM current is given
by (see Section \ref{sec:Main-calculation:-Fermi} below): 
\begin{equation}
I=I_{1}+I_{2}+I_{3}\label{eq:Current-1}
\end{equation}
Where:
\begin{align}
I_{1} & =-\frac{1}{\pi}e\int_{-\infty}^{\infty}d\omega\left[n_{F}\left(\omega-e\mathcal{V}\right)-n_{F}\left(\omega\right)\right]\times\nonumber \\
 & \qquad\qquad\times\left[A_{c_{\uparrow}^{\dagger}c_{\uparrow}}\left(\omega\right)+A_{c_{\downarrow}^{\dagger}c_{\downarrow}}\left(\omega\right)\right]\Gamma_{1}\label{eq:Current_1}
\end{align}
\begin{align}
I_{2} & =-\frac{1}{\pi}e\int_{-\infty}^{\infty}d\omega\left[n_{F}\left(\omega-e\mathcal{V}\right)-n_{F}\left(\omega\right)\right]\times\nonumber \\
 & \qquad\qquad\times2Re\left[A_{O_{c}^{\uparrow\dagger}c_{\uparrow}}\left(\omega\right)+A_{O_{c}^{\downarrow\dagger}c_{\downarrow}}\left(\omega\right)\right]\Gamma_{2}\label{eq:Current_2}
\end{align}

\begin{align}
I_{3} & =-\frac{1}{\pi}e\int_{-\infty}^{\infty}d\omega\left[n_{F}\left(\omega-e\mathcal{V}\right)-n_{F}\left(\omega\right)\right]\nonumber \\
 & \times\left[A_{O_{c}^{\uparrow\dagger}O_{c}^{\uparrow}}\left(\omega\right)+A_{O_{c}^{\downarrow\dagger}O_{c}^{\downarrow}}\left(\omega\right)\right]\Gamma_{3}\label{eq:Current_3}
\end{align}
where $\Gamma_{1/2/3}$ are related to the physics of the tip. This
is our main result which we derive in Section \ref{sec:Main-calculation:-Fermi}
below.

\section{\protect\label{sec:Main-calculation:-Fermi}Main calculation: Fermi
Golden rule for STM current in the presence of Coulomb assisted tunneling}

We will be closely following \citep{Coleman_2016}. Now we have that
the current is given by:
\begin{equation}
I=-e\left[P_{\psi\rightarrow c}-P_{c\rightarrow\psi}\right]\label{eq:Current}
\end{equation}
Where $P_{\psi\rightarrow c}$ is the probability of transition between
the tip and the sample and $P_{c\rightarrow\psi}$ is the probability
of transition between the sample and the tip. Now by Fermi's golden
rule we have that \citep{Shankar_1994,Coleman_2016}: 
\begin{align}
P_{\psi\rightarrow c} & =2\pi\sum_{\left|\lambda\right\rangle \left|\lambda'\right\rangle \left|\xi\right\rangle \left|\xi'\right\rangle }\delta\left(E_{\xi}+E_{\xi'}-E_{\lambda}-E_{\lambda'}\right)p_{\lambda}p_{\lambda'}\times\nonumber \\
 & \qquad\times\left|\left\langle \xi\right|\left\langle \xi'\right|\sum_{\sigma}\psi_{\sigma}^{\dagger}\left(0\right)c_{\sigma}\left(0\right)\times\right.\nonumber \\
 & \left.\times\left[t+\Omega n_{\psi\bar{\sigma}}\left(0\right)+\tilde{\Omega}n_{c\bar{\sigma}}\left(0\right)\right]\left|\lambda\right\rangle \left|\lambda'\right\rangle \right|^{2}\label{eq:Fermi_golden}
\end{align}
Where $\left|\lambda\right\rangle $ and $\left|\xi\right\rangle $
are exact eigenstates of $H_{c}$ while $\left|\lambda'\right\rangle $
and $\left|\xi'\right\rangle $ are exact eigenstates of $H_{\psi}$
and we have used Fermi's golden rule \citet{Coleman_2016,Shankar_1994}
and $\bar{\sigma}$is the opposite spin state of $\sigma$ and $p_{\lambda}$,
$p_{\lambda'}$ are Boltzmann weights. Therefore we have that:
\begin{align}
P_{\psi\rightarrow c} & =2\int_{-\infty}^{\infty}dt\sum_{\left|\lambda\right\rangle \left|\lambda'\right\rangle }p_{\lambda}p_{\lambda'}\left\langle \lambda\right|\left\langle \lambda'\right|\nonumber \\
 & \psi_{\uparrow}^{\dagger}\left(0\right)c_{\uparrow}\left(0\right)\left[t+\Omega n_{\psi\downarrow}\left(0\right)+\tilde{\Omega}n_{c\downarrow}\left(0\right)\right]\times\nonumber \\
 & \times\exp\left(iH_{\psi}t\right)\exp\left(iH_{c}t\right)\times\nonumber \\
 & \times\left[t^{*}+\Omega^{*}n_{\psi\downarrow}\left(0\right)+\tilde{\Omega}^{*}n_{c\downarrow}\left(0\right)\right]c_{\uparrow}^{\dagger}\left(0\right)\psi_{\uparrow}\left(0\right)\times\nonumber \\
 & \times\exp\left(-iH_{\psi}t\right)\exp\left(-iH_{c}t\right)\left|\lambda\right\rangle \left|\lambda'\right\rangle +\uparrow\Leftrightarrow\downarrow\label{eq:Probablity_direct}
\end{align}
 here we have Fourier transformed the delta function in Eq. (\ref{eq:Fermi_golden}).
This can further be simplified by going to the Heisenberg picture
\citep{Coleman_2016} as:
\begin{align}
P_{\psi\rightarrow c} & =2\int_{-\infty}^{-\infty}d\tau\sum_{\left|\lambda\right\rangle \left|\lambda'\right\rangle }p_{\lambda}p_{\lambda'}\left\langle \lambda\right|\left\langle \lambda'\right|\psi_{\uparrow}^{\dagger}\left(0\right)c_{\uparrow}\left(0\right)\times\nonumber \\
 & \times\left[t+\Omega n_{\psi\downarrow}\left(0\right)+\tilde{\Omega}n_{c\downarrow}\left(0\right)\right]\times\nonumber \\
 & \times\left[t^{*}+\Omega^{*}n_{\psi\downarrow}\left(\tau,0\right)+\tilde{\Omega}^{*}n_{c\downarrow}\left(\tau,0\right)\right]\times\nonumber \\
 & \times c_{\uparrow}^{\dagger}\left(\tau,0\right)\psi_{\uparrow}\left(\tau,0\right)\left|\lambda\right\rangle \left|\lambda'\right\rangle +\uparrow\Leftrightarrow\downarrow\label{eq:Heisenberg_picture}
\end{align}
We now expand all the terms to obtain $P_{\psi\rightarrow c}=$
\begin{align}
 & =2\int_{-\infty}^{\infty}d\tau\left[\left|t\right|^{2}\left\langle \psi_{\uparrow}^{\dagger}\left(0\right)\psi_{\uparrow}\left(\tau,0\right)\right\rangle \left\langle c_{\uparrow}\left(0\right)c_{\uparrow}^{\dagger}\left(\tau,0\right)\right\rangle \right.\nonumber \\
 & +t\Omega^{*}\left\langle n_{\psi\downarrow}\left(0\right)\psi_{\uparrow}^{\dagger}\left(0\right)\psi_{\uparrow}\left(\tau,0\right)\right\rangle \left\langle c_{\uparrow}\left(0\right)c_{\uparrow}^{\dagger}\left(\tau,0\right)\right\rangle \nonumber \\
 & +t^{*}\Omega\left\langle \psi_{\uparrow}^{\dagger}\left(0\right)\psi_{\uparrow}\left(\tau,0\right)n_{\psi\downarrow}\left(\tau,0\right)\right\rangle \left\langle c_{\uparrow}\left(0\right)c_{\uparrow}^{\dagger}\left(\tau,0\right)\right\rangle \nonumber \\
 & +t\tilde{\Omega}^{*}\left\langle \psi_{\uparrow}^{\dagger}\left(0\right)\psi_{\uparrow}\left(\tau,0\right)\right\rangle \left\langle c_{\uparrow}\left(0\right)c_{\uparrow}^{\dagger}\left(\tau,0\right)n_{c\downarrow}\left(\tau,0\right)\right\rangle \nonumber \\
 & +t^{*}\tilde{\Omega}\left\langle \psi_{\uparrow}^{\dagger}\left(0\right)\psi_{\uparrow}\left(\tau,0\right)\right\rangle \left\langle n_{c\downarrow}\left(0\right)c_{\uparrow}\left(0\right)c_{\uparrow}^{\dagger}\left(\tau,0\right)\right\rangle \nonumber \\
 & +\Omega\tilde{\Omega}^{*}\left\langle n_{\psi\downarrow}\left(0\right)\psi_{\uparrow}^{\dagger}\left(0\right)\psi_{\uparrow}\left(\tau,0\right)\right\rangle \left\langle c_{\uparrow}\left(0\right)c_{\uparrow}^{\dagger}\left(\tau,0\right)n_{c\downarrow}\left(\tau,0\right)\right\rangle \nonumber \\
 & +\Omega^{*}\tilde{\Omega}\left\langle \psi_{\uparrow}^{\dagger}\left(0\right)\psi_{\uparrow}\left(\tau,0\right)n_{\psi\downarrow}\left(\tau,0\right)\right\rangle \left\langle n_{c\downarrow}\left(\tau,0\right)c_{\uparrow}\left(0\right)c_{\uparrow}^{\dagger}\left(\tau,0\right)\right\rangle \nonumber \\
 & +\left|\Omega\right|^{2}\left\langle n_{\psi\downarrow}\left(0\right)\psi_{\uparrow}^{\dagger}\left(0\right)\psi_{\uparrow}\left(\tau,0\right)n_{\psi\downarrow}\left(\tau,0\right)\right\rangle \left\langle c_{\uparrow}\left(0\right)c_{\uparrow}^{\dagger}\left(\tau,0\right)\right\rangle \nonumber \\
 & \left.+\left|\tilde{\Omega}\right|^{2}\left\langle \psi_{\uparrow}^{\dagger}\left(0\right)\psi_{\uparrow}\left(\tau,0\right)\right\rangle \left\langle n_{c\downarrow}\left(0\right)c_{\uparrow}\left(0\right)c_{\uparrow}^{\dagger}\left(\tau,0\right)n_{c\downarrow}\left(\tau,0\right)\right\rangle \right]\nonumber \\
 & +\uparrow\Leftrightarrow\downarrow\label{eq:Simplified_expanded}
\end{align}
Where we have expanded all the terms in Eq. (\ref{eq:Heisenberg_picture}).
As such we have that \citep{Coleman_2016}: 
\begin{align}
P_{\psi\rightarrow c} & =2\int_{-\infty}^{\infty}d\tau\times\nonumber \\
 & \left[\left|t\right|^{2}\left\langle \psi_{\uparrow}^{\dagger}\left(0\right)\psi_{\uparrow}\left(\tau,0\right)\right\rangle \left\langle c_{\uparrow}\left(0\right)c_{\uparrow}^{\dagger}\left(\tau,0\right)\right\rangle +\right.\nonumber \\
 & +t\Omega^{*}\left\langle O_{\psi}^{\dagger}\psi_{\uparrow}\left(\tau,0\right)\right\rangle \left\langle c_{\uparrow}\left(0\right)c_{\uparrow}^{\dagger}\left(\tau,0\right)\right\rangle +\nonumber \\
 & +t^{*}\Omega\left\langle \psi_{\uparrow}^{\dagger}\left(0\right)O_{\psi}\left(\tau\right)\right\rangle \left\langle c_{\uparrow}\left(0\right)c_{\uparrow}^{\dagger}\left(\tau,0\right)\right\rangle +\nonumber \\
 & +t\tilde{\Omega}^{*}\left\langle \psi_{\uparrow}^{\dagger}\left(0\right)\psi_{\uparrow}\left(\tau,0\right)\right\rangle \left\langle c_{\uparrow}\left(0\right)O_{c}^{\dagger}\left(\tau\right)\right\rangle +\nonumber \\
 & +t^{*}\tilde{\Omega}\left\langle \psi_{\uparrow}^{\dagger}\left(0\right)\psi_{\uparrow}\left(\tau,0\right)\right\rangle \left\langle O_{c}c_{\uparrow}^{\dagger}\left(\tau,0\right)\right\rangle +\nonumber \\
 & +\Omega\tilde{\Omega}^{*}\left\langle O_{\psi}^{\dagger}\psi_{\uparrow}\left(\tau,0\right)\right\rangle \left\langle c_{\uparrow}\left(0\right)O_{c}^{\dagger}\left(\tau\right)\right\rangle +\nonumber \\
 & +\Omega^{*}\tilde{\Omega}\left\langle \psi_{\uparrow}^{\dagger}\left(0\right)O_{\psi}\right\rangle \left\langle O_{c}c_{\uparrow}^{\dagger}\left(\tau,0\right)\right\rangle +\nonumber \\
 & +\left|\Omega\right|^{2}\left\langle O_{\psi}^{\dagger}O_{\psi}\right\rangle \left\langle c_{\uparrow}\left(0\right)c_{\uparrow}^{\dagger}\left(\tau,0\right)\right\rangle \nonumber \\
 & \left.+\left|\tilde{\Omega}\right|^{2}\left\langle \psi_{\uparrow}^{\dagger}\left(0\right)\psi_{\uparrow}\left(\tau,0\right)\right\rangle \left\langle O_{c}O_{c}^{\dagger}\right\rangle \right]+\uparrow\Leftrightarrow\downarrow\label{eq:Transition_simplified}
\end{align}
Where we have used Eq. (\ref{eq:Operators}). To simplify the tedious
notation because we can focus only on one spin component (and there
is no spin orbit coupling) we can drop the spin index. As such:
\begin{align}
P_{\psi\rightarrow c} & =\frac{1}{\pi}\int_{-\infty}^{\infty}d\omega\left[\left|t\right|^{2}G_{\psi^{\dagger}\psi}^{<}\left(\omega\right)G_{c^{\dagger}c}^{>}\left(\omega\right)\right.\nonumber \\
 & +t\Omega^{*}G_{O^{\dagger}\psi}^{<}\left(\omega\right)G_{c^{\dagger}c}^{>}\left(\omega\right)+t^{*}\Omega G_{\psi^{\dagger}O}^{<}\left(\omega\right)G_{c^{\dagger}c}^{>}\left(\omega\right)\nonumber \\
 & +t\tilde{\Omega}^{*}G_{\psi^{\dagger}\psi}^{<}\left(\omega\right)G_{O^{\dagger}c}^{>}\left(\omega\right)+t^{*}\tilde{\Omega}G_{\psi^{\dagger}\psi}^{<}\left(\omega\right)G_{c^{\dagger}O}^{>}\left(\omega\right)\nonumber \\
 & +\Omega\tilde{\Omega}^{*}G_{O^{\dagger}\psi}^{<}\left(\omega\right)G_{O^{\dagger}c}^{>}\left(\omega\right)+\Omega^{*}\tilde{\Omega}G_{\psi^{\dagger}O}^{<}\left(\omega\right)G_{c^{\dagger}O}^{>}\left(\omega\right)\nonumber \\
 & \left.+\left|\Omega\right|^{2}G_{O^{\dagger}O}^{<}\left(\omega\right)G_{c^{\dagger}c}^{>}\left(\omega\right)+\left|\tilde{\Omega}\right|^{2}G_{\psi^{\dagger}\psi}^{<}\left(\omega\right)G_{O^{\dagger}O}^{>}\left(\omega\right)\right]\nonumber \\
 & +\uparrow\Leftrightarrow\downarrow\label{eq:Forward_current_frequency}
\end{align}
Now we use the relationship \citep{Coleman_2016}: 
\begin{align}
G^{>}\left(\omega\right) & =A\left(\omega\right)\left(1-n_{F}\left(\omega\right)\right)\nonumber \\
G^{<}\left(\omega\right) & =A\left(\omega\right)n_{F}\left(\omega\right)\label{eq:Lesser_greater}
\end{align}
Where $A\left(\omega\right)$ is defined by Eq. (\ref{eq:Greens_functions})
and $n_{F}$ is the Fermi function. As such we have that \citep{Coleman_2016}:
\begin{align}
P_{\psi\rightarrow c} & =\frac{1}{\pi}\int_{-\infty}^{\infty}d\omega\left[\left|t\right|^{2}A_{\psi^{\dagger}\psi}\left(\omega\right)A_{c^{\dagger}c}\left(\omega\right)\right.\nonumber \\
 & +t\Omega^{*}A_{O^{\dagger}\psi}\left(\omega\right)A_{c^{\dagger}c}\left(\omega\right)+t^{*}\Omega A_{\psi^{\dagger}O}\left(\omega\right)A_{c^{\dagger}c}\left(\omega\right)\nonumber \\
 & +t\tilde{\Omega}^{*}A_{\psi^{\dagger}\psi}\left(\omega\right)A_{O^{\dagger}c}\left(\omega\right)+t^{*}\tilde{\Omega}A_{\psi^{\dagger}\psi}\left(\omega\right)A_{c^{\dagger}O}\left(\omega\right)\nonumber \\
 & +\Omega\tilde{\Omega}^{*}A_{O^{\dagger}\psi}\left(\omega\right)A_{O^{\dagger}c}\left(\omega\right)+\Omega^{*}\tilde{\Omega}A_{\psi^{\dagger}O}\left(\omega\right)A_{c^{\dagger}O}\left(\omega\right)\nonumber \\
 & \left.+\left|\Omega\right|^{2}A_{O^{\dagger}O}\left(\omega\right)A_{c^{\dagger}c}\left(\omega\right)+\left|\tilde{\Omega}\right|^{2}A_{\psi^{\dagger}\psi}\left(\omega\right)A_{O^{\dagger}O}\left(\omega\right)\right]\times\nonumber \\
 & \times n_{F}\left(\omega-e\mathcal{V}\right)\left(1-n_{F}\left(\omega\right)\right)+\uparrow\Leftrightarrow\downarrow\label{eq:Forward_current}
\end{align}
Similarly we can now write \citep{Coleman_2016}:
\begin{align}
P_{c\rightarrow\psi} & =2\pi\sum_{\left|\lambda\right\rangle \left|\lambda'\right\rangle \left|\xi\right\rangle \left|\xi'\right\rangle }\delta\left(E_{\xi}+E_{\xi'}-E_{\lambda}-E_{\lambda'}\right)p_{\lambda}p_{\lambda'}\times\nonumber \\
 & \times\left|\left\langle \xi\right|\left\langle \xi'\right|\sum_{\sigma}\left[t^{*}+\Omega^{*}n_{\psi\downarrow}\left(0\right)+\tilde{\Omega}^{*}n_{c\downarrow}\left(0\right)\right]\times\right.\nonumber \\
 & \left.\times c_{\sigma}^{\dagger}\left(0\right)\psi_{\sigma}\left(0\right)\left|\lambda\right\rangle \left|\lambda'\right\rangle \right|^{2}\label{eq:Back_flow}
\end{align}
As such \citep{Coleman_2016}:
\begin{align}
P_{\psi\rightarrow c} & =\frac{1}{\pi}\int_{-\infty}^{\infty}d\omega\left[\left|t\right|^{2}A_{\psi^{\dagger}\psi}\left(\omega\right)A_{c^{\dagger}c}\left(\omega\right)\right.\nonumber \\
 & +t\Omega^{*}A_{O^{\dagger}\psi}\left(\omega\right)A_{c^{\dagger}c}\left(\omega\right)+t^{*}\Omega A_{\psi^{\dagger}O}\left(\omega\right)A_{c^{\dagger}c}\left(\omega\right)\nonumber \\
 & +t\tilde{\Omega}^{*}A_{\psi^{\dagger}\psi}\left(\omega\right)A_{O^{\dagger}c}\left(\omega\right)+t^{*}\tilde{\Omega}A_{\psi^{\dagger}\psi}\left(\omega\right)A_{c^{\dagger}O}\left(\omega\right)\nonumber \\
 & +\Omega\tilde{\Omega}^{*}A_{O^{\dagger}\psi}\left(\omega\right)A_{O^{\dagger}c}\left(\omega\right)+\Omega^{*}\tilde{\Omega}A_{\psi^{\dagger}O}\left(\omega\right)A_{c^{\dagger}O}\left(\omega\right)\nonumber \\
 & \left.+\left|\Omega\right|^{2}A_{O^{\dagger}O}\left(\omega\right)A_{c^{\dagger}c}\left(\omega\right)+\left|\tilde{\Omega}\right|^{2}A_{\psi^{\dagger}\psi}\left(\omega\right)A_{O^{\dagger}O}\left(\omega\right)\right]\times\nonumber \\
 & \times n_{F}\left(\omega\right)\left(1-n_{F}\left(\omega-e\mathcal{V}\right)\right)+\uparrow\Leftrightarrow\downarrow\label{eq:Backward_current}
\end{align}
This means the current can be rewritten as:
\begin{widetext}
\begin{align}
I= & -\frac{1}{\pi}e\sum_{\sigma}\int_{-\infty}^{\infty}d\omega\left[n_{F}\left(\omega-e\mathcal{V}\right)-n_{F}\left(\omega\right)\right]\times A_{c_{\sigma}^{\dagger}c_{\sigma}}\left(\omega\right)\left[\left|t\right|^{2}A_{\psi_{\bar{\sigma}}^{\dagger}\psi_{\bar{\sigma}}}\left(\omega\right)+2Re\left[t\Omega^{*}A_{O_{\psi}^{\bar{\sigma}\dagger}\psi_{\bar{\sigma}}}\left(\omega\right)\right]+\left|\Omega\right|^{2}A_{O_{\psi}^{\bar{\sigma}\dagger}O_{\psi}^{\bar{\sigma}}}\left(\omega\right)\right]\nonumber \\
 & -\frac{1}{\pi}e\sum\int_{\sigma-\infty}^{\infty}d\omega\left[n_{F}\left(\omega-e\mathcal{V}\right)-n_{F}\left(\omega\right)\right]\times2Re\left[A_{O_{c}^{\sigma\dagger}c_{\sigma}}\left(\omega\right)\left[t\tilde{\Omega}^{*}A_{\psi_{\bar{\sigma}}^{\dagger}\psi_{\bar{\sigma}}}\left(\omega\right)+\Omega\tilde{\Omega}^{*}A_{O_{\psi}^{\bar{\sigma}\dagger}\psi_{\bar{\sigma}}}\left(\omega\right)\right]\right]\nonumber \\
 & -\frac{1}{\pi}e\sum\int_{\sigma-\infty}^{\infty}d\omega\left[n_{F}\left(\omega-e\mathcal{V}\right)-n_{F}\left(\omega\right)\right]\times A_{O_{c}^{\dagger}O_{c}}\left(\omega\right)\left[\left|\tilde{\Omega}\right|^{2}A_{\psi^{\dagger}\psi}\left(\omega\right)\right]\label{eq:Current_transparent}
\end{align}
\end{widetext}

Now we work in the broad tip limit whereby we obtain Eq. (\ref{eq:Current-1}).

\section{\protect\label{sec:Calculation-for-free}Calculation for free fermions
in the uniform wide band limit}

We introduce free fermions with a band with $\left[-D,D\right]$,
a density $\rho$and a chemical potential $\mu$ and we will work
in the zero temperature limit. Then we have that: 
\begin{equation}
A_{c^{\dagger}c}\left(\omega\right)=\rho\theta\left(D-\omega\right)\theta\left(\omega+D\right)\label{eq:Known}
\end{equation}
Furthermore 
\begin{equation}
A_{O_{c}^{\dagger}c}\left(\omega\right)=2\rho\left(D+\mu\right)\rho\theta\left(D-\omega\right)\theta\left(\omega+D\right)\label{eq:Simple}
\end{equation}
Now for free fermions 
\begin{equation}
A_{O_{c}^{\dagger}O_{c}}\left(\omega\right)=-\pi\rho^{3}\omega^{2}\theta\left(-\omega\right)\label{eq:Calculate}
\end{equation}
See Appendix \ref{sec:Free-fermions-in}. Here we have dropped the
spin index as its just a facto of two. This means that: 
\begin{equation}
I=I_{1}+I_{2}+I_{3}\label{eq:Current_final}
\end{equation}
Where: 
\begin{align}
I_{1} & =-\frac{2}{\pi}e\int_{-\infty}^{\infty}d\omega\left[n_{F}\left(\omega-e\mathcal{V}\right)-n_{F}\left(\omega\right)\right]\times\nonumber \\
 & \qquad\qquad\times\rho\theta\left(D-\omega\right)\theta\left(\omega+D\right)\Gamma_{1}\label{eq:Current_1-1}
\end{align}
\begin{align}
I_{2} & =-\frac{8}{\pi}e\int_{-\infty}^{\infty}d\omega\left[n_{F}\left(\omega-e\mathcal{V}\right)-n_{F}\left(\omega\right)\right]\times\nonumber \\
 & \qquad\qquad\times\rho^{2}\left(D+\mu\right)\theta\left(D-\omega\right)\theta\left(\omega+D\right)\Gamma_{2}\label{eq:Current_2-1}
\end{align}
\begin{align}
I_{3} & =+2e\int_{-\infty}^{\infty}d\omega\left[n_{F}\left(\omega-e\mathcal{V}\right)-n_{F}\left(\omega\right)\right]\times\nonumber \\
 & \qquad\qquad\times\rho^{3}\omega^{2}\theta\left(-\omega\right)\Gamma_{3}\theta\left[D-\omega\right]\theta\left[D+\omega\right]\label{eq:Current_3-1}
\end{align}

\section{\protect\label{sec:Conclusions}Conclusions}

In this work we have extended the calculation of STM current to the
case of Coulomb assisted tunneling between the tip and the sample
using the extended two site Hubbard model analysis for the atom in
the tip and the atom in the sample closest to each other. In this
case on top of the usual tunneling, which comes about when electrons
can tunnel through the classically forbidden region between the tip
and the sample, we also consider the case when there is a Coulomb
interaction, as in the extended Hubbard model, that helps the tunneling
\citep{Fazekas_1999} and introduces additional terms to the Hamiltonian
(Coulomb assisted tunneling \citep{Fazekas_1999,Hubbard_1963,Hubbard_1964,Hubbard_1965}).
In this case there are additional terms for interactions between the
tip and the sample which lead to additional currents. We have computed
these terms and currents and shown they have non-trivial consequences
even in the case of free fermions with a uniform density of states
where the $\frac{\partial I}{\partial\mathcal{V}}$ curve now depends
on the voltage $\mathcal{V}$. We have shown that the correction terms
are order one and for simplicity here we have ignored spin orbit coupling.
In the future it would be of interest to apply these results to real
experiments.

\appendix

\section{\protect\label{sec:Free-fermions-in}Free fermions in the wide band
limit}

We now compute for free fermions in the wide band limit:
\begin{align}
 & \left\langle n_{\uparrow}\left(0\right)n_{\uparrow}\left(\tau\right)\right\rangle =\rho^{2}\int_{-D}^{D}d\varepsilon_{1}\int_{-D}^{D}d\varepsilon_{2}\times\nonumber \\
 & \times f\left[\varepsilon_{1}-\mu\right]\left[1-f\left[\varepsilon_{2}-\mu\right]\right]\exp\left[i\left[\varepsilon_{2}-\varepsilon_{1}\right]\tau\right]\nonumber \\
 & \left\langle c_{\downarrow}^{\dagger}\left(0\right)c_{\downarrow}\left(\tau\right)\right\rangle \nonumber \\
 & =\rho\int_{-D}^{D}d\varepsilon_{3}f\left[\varepsilon_{3}-\mu\right]\exp\left[-i\left[\varepsilon_{3}-\mu\right]\tau\right]\label{eq:Free_fermions-2}
\end{align}
and similarly for spin down and the times reversed. This means that: 
\begin{widetext}
\begin{align}
A_{O_{c}^{\uparrow\dagger}O_{c}^{\uparrow}}\left(\omega\right) & =-\rho^{3}\int_{-D}^{D}d\varepsilon_{1}\int_{-D}^{D}d\varepsilon_{2}\int_{-D}^{D}d\varepsilon_{3}\int d\omega\exp\left(-i\left(\varepsilon_{1}-\varepsilon_{2}+\left(\varepsilon_{3}-\mu\right)-\omega\right)\tau\right)f\left(\varepsilon_{1}-\mu\right)\left(1-f\left(\varepsilon_{2}-\mu\right)\right)f\left(\varepsilon_{3}-\mu\right)\nonumber \\
 & +\rho^{3}\int_{-D}^{D}d\varepsilon_{1}\int_{-D}^{D}d\varepsilon_{2}\int_{-D}^{D}d\varepsilon_{3}\int d\omega\exp\left(i\left(\varepsilon_{1}-\varepsilon_{2}+\left(\varepsilon_{3}-\mu\right)-\omega\right)\tau\right)f\left(\varepsilon_{1}-\mu\right)\left(1-f\left(\varepsilon_{2}-\mu\right)\right)\left[1-f\left(\varepsilon_{3}-\mu\right)\right]\label{eq:Spectral}
\end{align}
Then we obtain: 
\begin{align}
A_{O_{c}^{\uparrow\dagger}O_{c}^{\uparrow}}\left(\omega\right) & =-2\pi\rho^{3}\int_{-D}^{D}d\varepsilon_{1}\int_{-D}^{D}d\varepsilon_{2}\int_{-D}^{D}d\varepsilon_{3}f\left(\varepsilon_{1}-\mu\right)\left(1-f\left(\varepsilon_{2}-\mu\right)\right)f\left(\varepsilon_{3}-\mu\right)\delta\left(\omega+\varepsilon_{1}-\varepsilon_{2}+\varepsilon_{3}-\mu\right)\nonumber \\
 & +2\pi\rho^{3}\int_{-D}^{D}d\varepsilon_{1}\int_{-D}^{D}d\varepsilon_{2}\int_{-D}^{D}d\varepsilon_{3}f\left(\varepsilon_{1}-\mu\right)\left(1-f\left(\varepsilon_{2}-\mu\right)\right)\left[1-f\left(\varepsilon_{3}-\mu\right)\right]\delta\left(\omega-\left[\varepsilon_{1}-\varepsilon_{2}+\varepsilon_{3}-\mu\right]\right)\label{eq:Spectral_II}
\end{align}

Continuing: 
\begin{align}
A_{O_{c}^{\uparrow\dagger}O_{c}^{\uparrow}}\left(\omega\right) & =-2\pi\rho^{3}\int_{\mu}^{D}d\varepsilon_{1}\int_{-D}^{\mu}d\varepsilon_{2}\int_{-D}^{\mu}d\varepsilon_{3}\delta\left(\omega+\left[\varepsilon_{1}-\varepsilon_{2}+\varepsilon_{3}-\mu\right]\right)+\nonumber \\
 & \qquad+2\pi\rho^{3}\int_{\mu}^{D}d\varepsilon_{1}\int_{-D}^{\mu}d\varepsilon_{2}\int_{\mu}^{D}d\varepsilon_{3}\delta\left(\omega-\left[\varepsilon_{1}-\varepsilon_{2}+\varepsilon_{3}-\mu\right]\right)\label{eq:Spectral_III}
\end{align}

\begin{align}
A_{O_{c}^{\uparrow\dagger}O_{c}^{\uparrow}}\left(\omega\right) & =-2\pi\rho^{3}\int_{\mu}^{D}d\varepsilon_{1}\int_{-D}^{\mu}d\varepsilon_{2}\theta\left(-\omega-\left[\varepsilon_{1}-\varepsilon_{2}\right]\right)\theta\left(\omega+D+\left[\varepsilon_{1}-\varepsilon_{2}-\mu\right]\right)+\nonumber \\
 & \qquad+2\pi\rho^{3}\int_{\mu}^{D}d\varepsilon_{1}\int_{-D}^{\mu}d\varepsilon_{2}\theta\left(\omega-\left[\varepsilon_{1}-\varepsilon_{2}\right]\right)\theta\left(-\omega-D+\left[\varepsilon_{1}-\varepsilon_{2}-\mu\right]\right)\label{eq:Spectral_IV}
\end{align}

\begin{align}
A_{O_{c}^{\uparrow\dagger}O_{c}^{\uparrow}}\left(\omega\right) & =2\pi\rho^{3}\left[\int_{0}^{D-\mu}d\varepsilon_{1}\int_{0}^{D+\mu}d\varepsilon_{2}\left[\theta\left(\omega-\left[\varepsilon_{1}+\varepsilon_{2}\right]\right)\theta\left(-\omega-D-\mu+\left[\varepsilon_{1}+\varepsilon_{2}\right]\right)\right]\right.\nonumber \\
 & \qquad\left.-\theta\left(-\omega-\left[\varepsilon_{1}+\varepsilon_{2}\right]\right)\theta\left(\omega+D-\mu+\left[\varepsilon_{1}+\varepsilon_{2}\right]\right)\right]\label{eq:Spectral_V}
\end{align}

\begin{align}
A_{O_{c}^{\uparrow\dagger}O_{c}^{\uparrow}}\left(\omega\right) & \cong-2\pi\rho^{3}\left[\int_{0}^{D-\mu}d\varepsilon_{1}\int_{0}^{D+\mu}d\varepsilon_{2}\theta\left(-\omega-\left[\varepsilon_{1}+\varepsilon_{2}\right]\right)\theta\left(D-\left[\varepsilon_{1}+\varepsilon_{2}\right]\right)\right]\theta\left[D-\omega\right]\theta\left[D+\omega\right]\nonumber \\
 & =-2\pi\rho^{3}\left[\int_{0}^{2D}d\mathcal{E}\int_{-\mathcal{E}/2}^{\mathcal{E}/2}d\varepsilon\theta\left(-\omega-\mathcal{E}\right)\theta\left(D-\mathcal{E}\right)\right]\theta\left[D-\omega\right]\theta\left[D+\omega\right]\label{eq:Spectral_VI}
\end{align}
\end{widetext}

Combining we get that: 
\begin{align}
A_{O_{c}^{\uparrow\dagger}O_{c}^{\uparrow}}\left(\omega\right) & =-2\pi\rho^{3}\theta\left[D-\omega\right]\theta\left[D+\omega\right]\int_{0}^{D}d\mathcal{E}\mathcal{E}\theta\left(-\omega-\mathcal{E}\right)\nonumber \\
 & =-\pi\rho^{3}\omega^{2}\theta\left(-\omega\right)\theta\left[D-\omega\right]\theta\left[D+\omega\right]\label{eq:Spectral_VII}
\end{align}

The calculation for the spin down spectral function is identical.

\end{document}